\documentclass[9pt,twocolumn,twoside]{pnas-new}

\articletype{CLASSIFICATION}

\templatetype{pnasresearcharticle} 

\begin{document}

\title{Wind farms as sensor arrays of turbulent boundary layer spatio-temporal flow structure}

\author[a,1]{Manuel Ayala}
\author[a,1]{Dennice F. Gayme}
\author[a,1,2]{Charles Meneveau}

\affil[a]{Department of Mechanical Engineering, Johns Hopkins University, Baltimore, MD 21218}

\leadauthor{Ayala}

\significancestatement{

Wind farms supply an increasing
share of the world’s electricity, but the
temporal variability of the supplied
power poses inherent challenges,
because fluctuations can affect grid
balancing, and reliability.
There is great interest in predicting the fluctuation’s statistical features  
to aid in wind farm design
and operation, yet predictions are difficult
without detailed simulations
or site-specific measurements. We
show that established knowledge about turbulent 
boundary layer physics can be used
to predict the frequency spectrum of
wind-farm power fluctuations directly
from atmospheric conditions, turbine
properties, and wind farm layout. The
model illustrates how fundamental
fluid mechanics and turbulence phenomenological theories
can provide practical, a priori
information critical to integrating an
inherently variable renewable energy
source into power systems
}

\authorcontributions{M.A, D.F.G and C.M. designed research; M.A. performed research; M.A.,
D.F.G., and C.M. analyzed data; and M.A., D.F.G., and C.M. wrote the paper.}
\authordeclaration{The authors declare no competing interest.}
\correspondingauthor{\textsuperscript{2}To whom correspondence should be addressed. E-mail: meneveau@jhu.edu}


\begin{abstract}
Temporal fluctuations in wind-farm power arise from interactions between atmospheric turbulence, turbine response, and array geometry. These fluctuations affect turbine loading, farm-level power quality, and ancillary-service requirements, yet their accurate prediction remains challenging. Here, we develop an analytical framework for predicting the temporal spectrum of wind-farm power fluctuations and evaluate it against large-eddy simulations of a wind farm operating within a conventionally neutral atmospheric boundary layer. The framework combines three established elements of turbulent boundary-layer physics: a spatio-temporal turbulence spectrum that accounts for mean advection and random sweeping by large eddies; a top-down model of the fully developed wind-farm boundary layer that provides the required mean-flow and turbulence scales; and a spatial sampling kernel that represents turbine locations and finite rotor extent, also capturing
vertical filtering of turbulent fluctuations across the rotor. 
Using only atmospheric, turbine, and layout parameters, the model accurately predicts the aggregate power spectrum as measured from high-fidelity, large-eddy-simulation data of flow over a wind farm in flat terrain. The model reproduces peaks associated with advection between turbine rows, the attenuation of inertial-range fluctuations by rotor averaging, and spectra for multiple turbine arrangements, including staggered and randomly selected clusters. The framework also clarifies how convection velocity, turbulence intensity, rotor size, and turbine spacing independently control spectral peak locations, amplitudes, and broadening across conditions. These results show that established fluid-dynamic theory can provide fully a priori predictions of wind-farm power variability and may support improved wind-farm design, grid integration, and assessment of balancing and reserve requirements.

\end{abstract}

\dates{This manuscript was compiled on \today}

\maketitle
\thispagestyle{firststyle}
\ifthenelse{\boolean{shortarticle}}{\ifthenelse{\boolean{singlecolumn}}{\abscontentformatted}{\abscontent}}{}

\Firstpage

\begin{figure*}[t!]
\centering
\includegraphics[width=1.04\linewidth]{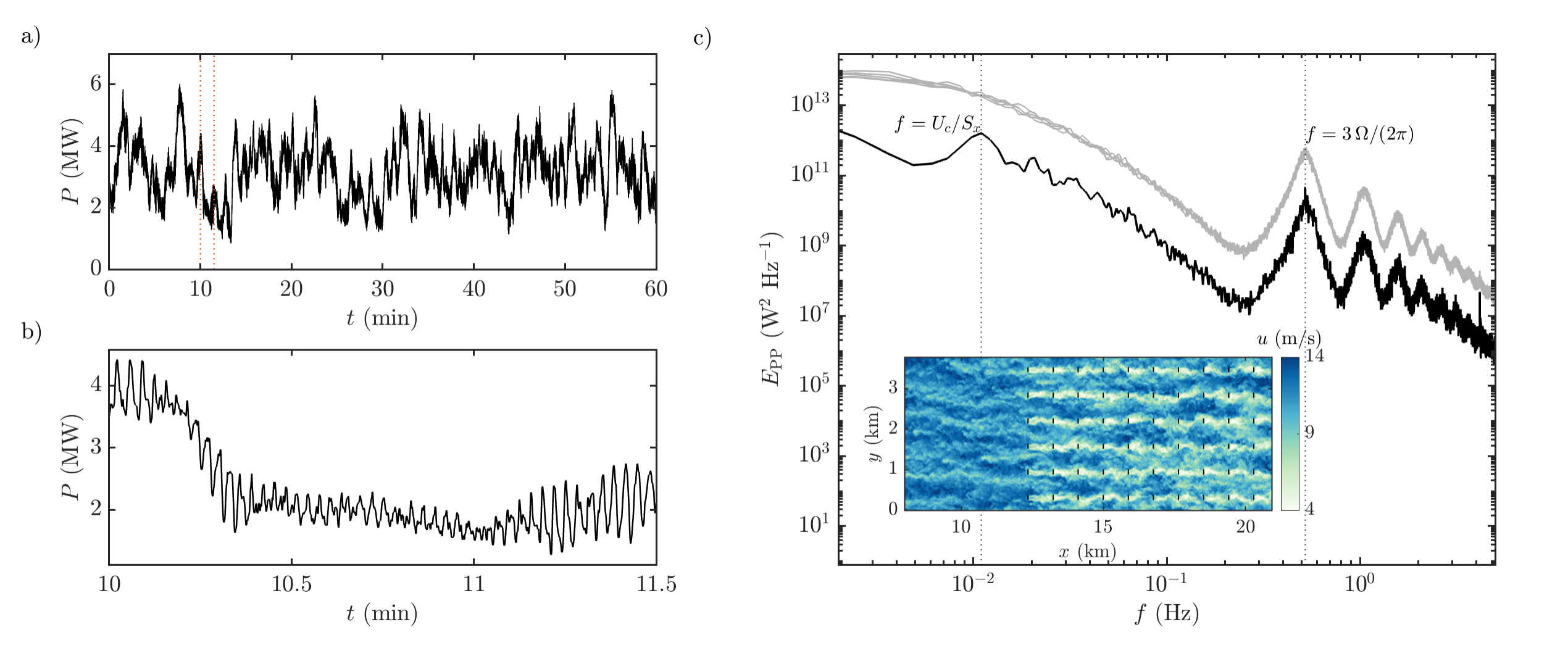}
\caption{(a) Aerodynamic torque-based power-output 1-hr time series for a representative NREL 5-MW turbine located within the fully developed region of  a wind farm modeled using high-fidelity numerical simulation. (b) 90 seconds subset in which blade rotation periodicity can be observed. (c) Power spectral density of the aggregate power fluctuations from the 30 turbines comprising the final five rows of the wind farm (black). Grey lines show the single-turbine power spectra, each averaged over the six turbines within one of the five rows. $U_c=9.7$m/s is the convective velocity of the flow over the turbine boundary layer, $S_x=882$m is the streamwise spacing between turbine rows and $\Omega= 1.09$(rad/s)  is rotational speed of speed of the turbine blades. Inset shows the contour of the streamwise velocity field at hub height level for a region of the domain at t = 1000 sec.}
\label{fig:fig1}
\end{figure*}

One of the major challenges associated with wind farm-generated power is its inherent variability over many temporal and spatial scales \cite{apt2007,bandi2017}. Power fluctuations affect turbine loading, and electricity grid integration requires compensation for power fluctuations through e.g., control strategies \cite{Shapiro2022TurbulenceControl}, storage or other power sources \cite{KatzensteinApt2012}. Therefore, accurate characterization of wind generated power fluctuations is critical for efficient grid operations under scenarios of increased wind energy penetration. The spectral density of temporal power fluctuations aggregated across several turbines was considered in pioneering work by \cite{apt2007} who observed a power-law dependence with a Kolmogorov-like $f^{-5/3}$ scaling spanning several orders of magnitude in frequency. This result prompted questions about how the power spectrum relates to the scaling of velocity fluctuations in the turbulent atmospheric boundary layer \cite{Stevens2014Temporal,milan_turbulent_2013,bandi2017,Bossuyt2017,chamorro_turbulence_2015,Tobin2015}. At time-scales in a range between seconds and tens of minutes, the aggregate wind farm power fluctuations depend crucially upon the spatio-temporal statistics and structures of fluid turbulence in the incoming wind \cite{Bossuyt2017}. 
Conversely, the variability of the incoming wind velocity at long time-scales (e.g. above 30 minutes to hours) is caused by passage of large-scale weather patterns and the diurnal cycle, and is therefore outside the realm of 
\Endparasplit
fluid turbulence. Fluctuations on the order of seconds and faster are irrelevant for large turbines whose response time is slower. 
In this work, we focus on the range of time-scales for which fluid turbulence is directly relevant, namely between seconds and tens of minutes. As an example,  Figure~\ref{fig:fig1}(a,b) shows the aerodynamic power, defined as turbine torque times rotor speed, for a representative turbine in a high-fidelity simulation of a 60-turbine wind farm under conventionally neutral conditions \cite{Zhu2025JHTDBWind}. Panel~(c) compares its power spectrum with that of the aggregate signal from 30 turbines in the fully developed region of the farm. The aggregate spectrum exhibits a peak near $10^{-2}$~Hz associated with advection between turbine rows and peaks near $0.5$~Hz and its harmonics due to rotational blade sampling. Our objective is to predict the spectral decay between $0.002$ and $0.2$~Hz from turbine properties, farm layout, and boundary-layer turbulence.

Considerable research has been devoted over the past decades to elucidating and characterizing the velocity fluctuations in turbulent boundary layers \cite{Pope2000Turbulent,Marusic2010HighReynoldsPoF,Jimenez2018Coherent}. Data from laboratory and field experiments \cite{Smits2011HighReynolds,Hutchins2012Reconciling}, as well as numerical simulations \cite{Lee_Moser_2015,Kim1987ChannelFlow,Sillero2013OnePoint} and phenomenological models \cite{marusic_review_AEM,Townsend1976TurbulentShearFlow} have contributed greatly to our understanding of wall-bounded turbulence and associated fluid mechanical phenomena. In particular, the distribution of mean velocity and variance of velocity fluctuations as function of distance to the bounding surface (the wall or the ground in the case of the atmospheric boundary layer) has been studied in depth \cite{Marusic2013LogarithmicRegion,Vallikivi2015HighReynolds,Hultmark2013LogarithmicScaling}. Progress has been made in understanding the spatial structure and correlations of the velocity fluctuations between two points in the boundary layer \cite{Wallace2014SpaceTime,Ganapathisubramani2005LargeScale}. Beyond purely temporal correlations \cite{Quadrio2003Integral}, substantial progress has been made in characterizing the full spatio-temporal structure of boundary-layer turbulence \cite{Wilczek_Stevens_Meneveau_2015,wilczek_height-dependence_2015,He_spacetime_review}.

An important question is the extent to which, such fundamental knowledge about boundary layer fluid mechanics and turbulence shed light on the problem of wind power variability? Treating turbines as discrete samplers of turbulent velocity fluctuations, \cite{Bossuyt2017} developed a simplified model for wind-farm power spectra and compared it with wind-tunnel measurements using porous-disk turbine surrogates. However, the model neglected vertical turbulence variations, and the validation data did not include rotating blades.

The purpose of this study is to develop a more complete spectral model for wind-power fluctuations and evaluate it against high-fidelity simulations from the recently developed JHTDB-wind database, which includes the effects of rotating blades \cite{Zhu2025JHTDBWind}. The framework combines established models for the mean velocity and turbulence variance in wall-bounded flows \cite{smits_annualreview_2010}, the Kraichnan random-sweeping hypothesis for spatio-temporal turbulence statistics \cite{Kraichnan1964Kolmogorov,Wilczek_narita_2012,Wilczek_Stevens_Meneveau_2015,wilczek_height-dependence_2015}, Kolmogorov scaling for the small-scale spectrum \cite{Kolmogorov1941Local}, and Townsend’s equilibrium-range hypothesis for larger scales \cite{Townsend1976TurbulentShearFlow}. The mean flow within the wind farm is predicted using the top-down model \cite{Frandsen1992WindSpeedReduction,calaf2010large,Meneveau2012TopDown}.

\section*{Modeling the wind-farm power-output spectrum}
\label{sec:section1}

\subsection*{From turbulent velocity fluctuations to power fluctuations} We first establish the connection between fluctuations in the turbulent velocity field and fluctuations in wind-turbine power. For turbine $i$, let $\langle u \rangle_{D,i}(t)$ denote the incoming streamwise velocity spatially averaged over the rotor disk. On time scales longer than the blade-rotation time, the instantaneous aerodynamic power can be approximated using the quasi-steady actuator-disk relation

\begin{equation}
    P_i(t)
    =
    \frac{1}{2}\rho \, A \, C'_P \,
    \langle u \rangle_{D,i}^{\,3}(t),
    \label{eqn:diskavg_power}
\end{equation}
where $\rho$ is the air density, $A=\pi D^2/4$ is the rotor area with diameter $D$, and
$C'_P=C_P/(1-a)^3$ is the power coefficient based on the disk-averaged velocity, with $a$ denoting the induction factor. We focus on turbines operating in  ``region 2'' (where inlet velocities are above the turbine cut-in speed but below-rated). Under these conditions, the turbine controller seeks to maintain near-optimal aerodynamic efficiency, and $C'_P$ can be considered constant in time.

Decomposing the disk-averaged velocity into a temporal mean $\langle \overline{u} \rangle_{D,i}$ and a temporally fluctuating component $\langle u \rangle^\prime_{D,i}$, i.e. $ \langle u \rangle_{D,i} =  \langle \overline{u} \rangle_{D,i} + \langle u \rangle^\prime_{D,i}$ and linearizing
Eq.~\ref{eqn:diskavg_power} about the temporal mean (assuming that 
$\langle u \rangle^\prime_{D,i} << \langle \overline{u} \rangle_{D,i}$)
yields  power fluctuations \cite{Bossuyt2017}: 
\begin{equation}
    P'_i(t)
    \approx
    C_{2}\,\langle u \rangle'_{D,i}(t),
    \qquad
    C_{2}
    =
    \frac{3}{2}\rho A C'_P\langle \overline{u} \rangle_{D}^{\,2},
    \label{eqn:power2fluc}
\end{equation}
where we assume that the mean velocity is the same for all turbines, which is appropriate within the fully developed region of a large wind farm. Thus, to leading order, fluctuations in turbine power are proportional to fluctuations in the disk-averaged incoming velocity. Moreover, higher powers of $u'$ scale similarly \cite{bandi2017}, providing additional support for this approximation.

The fluctuating power output of a wind farm containing $N$ turbines is given by the sum of the individual turbine contributions,
$P'_{\mathrm{wf}}(t)=\sum P'_i(t)\approx C_{2}\sum \langle u' \rangle_{D,i}(t)$. This relation supports the notion that the aggregate wind-farm power can be considered as a discrete sampling (summation) combined with a  spatial filtering of the turbulent velocity field over the rotor disk \cite{Bossuyt2017}. Introducing a sampling kernel $G(\boldsymbol{x})$, which accounts for both the sampling at turbine locations and the spatial averaging performed by each rotor, the wind-farm power fluctuations can be written as 
\begin{equation}
P'_{\mathrm{wf}}(t)\approx C_2 \iiint G(\boldsymbol{x})\, u'(\boldsymbol{x},t)\,\mathrm{d}\boldsymbol{x}.
\label{eq:linearrel}
\end{equation}
The turbine array therefore does not simply average a collection of independent power signals, but a spatially correlated turbulent field at a prescribed set of locations. The magnitude and time scales of the aggregate power fluctuations consequently depend on both the spatio-temporal structure of the turbulence and the geometry of the wind farm. 

The linear relation between power and turbulent velocity fluctuations (Eq. \ref{eq:linearrel}) and Parseval's theorem relating integrals in ${\bf x}$ and ${\bf k}$ in 3D imply that the power-spectral density $E_{\text{PP}}(\omega)$ (e.g. as shown in Fig. \ref{fig:fig1}) can be written in terms of the fluctuating turbulent velocity wave-number-frequency spectrum 
$E_{uu}({\bf k},\omega)$ according to

\begin{equation}
    E_{\text{PP}}(\omega) = C_2^2 \iiint |\hat{G}(\boldsymbol{k})|^2\,E_{\text{uu}}(\boldsymbol{k},\omega)\, d^3\boldsymbol{k},
    \label{eqn:E_pp}
\end{equation}
where $\boldsymbol{k}=(k_x,k_y,k_z)^{\mathrm{T}}$ is the three-dimensional wave vector, and $\omega$ is the temporal frequency. 

\subsection*{Three-dimensional spatial sampling by the turbine array} We next construct the spatial sampling function $G(\boldsymbol{x})$, which encodes the turbine positions, wind-farm layout, and spatial filtering introduced by the rotor disks. 


For turbines located at a common hub height $z_h$, we write the three-dimensional sampling function as $G(\boldsymbol{x}) =G_z(z)\,G_{xy}(x,y),$ where $G_{xy}$ describes the horizontal distribution of the turbines and $G_z$ represents the vertical filtering associated with the rotor area. The
horizontal sampling function is defined as
\begin{equation}
 G_{xy}(x,y) = \sum_{i=1}^N\delta(x-x_i)\,\,\frac{1}{D} \text{H} \left(\frac{D}{2} - |y-y_i| \right),
 \label{eq:G_xy}
\end{equation}
where $(x_i,y_i)$ denotes the horizontal position of turbine $i$,
$\delta(\cdot)$ is the Dirac delta function, and $\mathrm{H}(\cdot)$ is the Heaviside function. The Dirac delta function prescribes the discrete streamwise sampling by the turbine array, and the box filter accounts for the averaging across the spanwise extent of each rotor.
\begin{figure*}[b!]
\centering
\includegraphics[width=\linewidth]{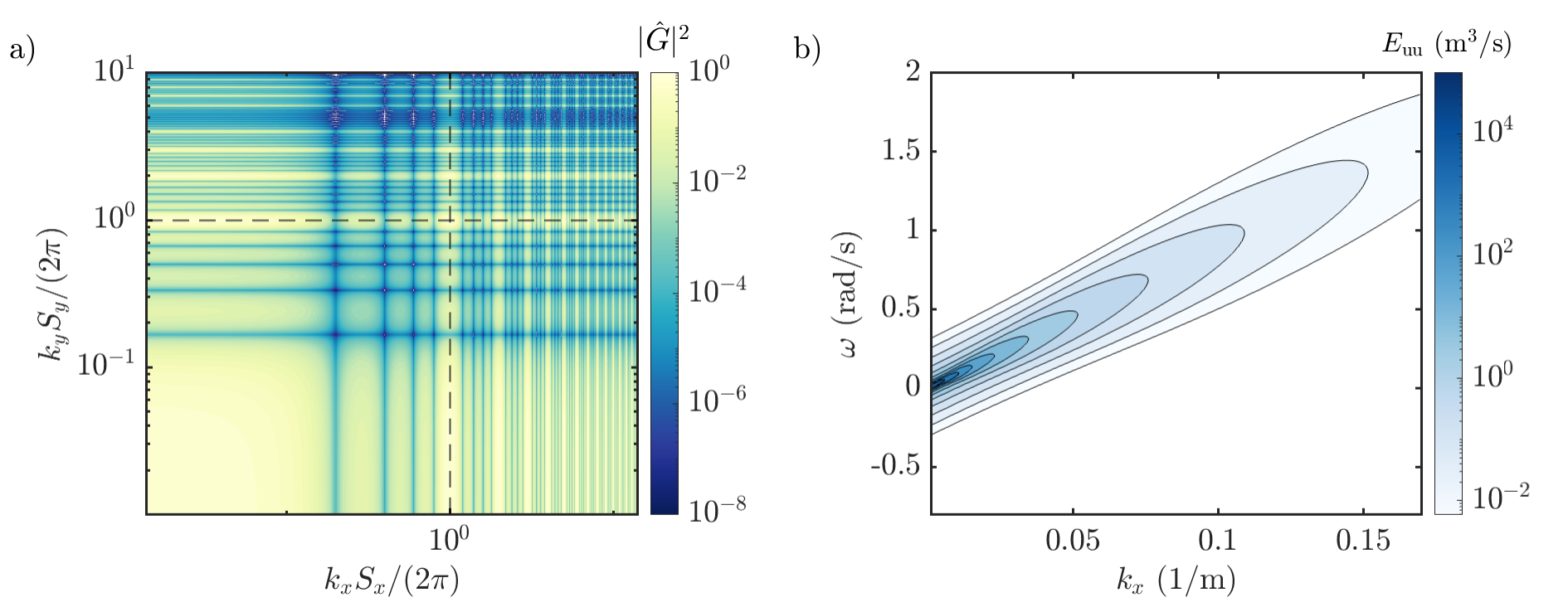}
\caption{a) Spectral transfer function (Eq.~\ref{eq:G_transfer}) representing the three-dimensional spatial sampling of a $5\times6$ wind farm array. The transfer function $\hat{G}(\boldsymbol{k})$ is divided by the number of turbines N to to represent the relative reduction of the spectra.  (b) Spatio-temporal spectra of the streamwise fluctuations estimated by Eqs.~\ref{eq:LRA_model} - \ref{eq:E_large_2D}.}
\label{fig:fig2}
\end{figure*}

To incorporate the vertical extent of the rotor, we use a spectral cutoff-filter (for subsequent analytical convenience) with kernel
\begin{equation}
    G_z(z) = 
    \frac{1}{D}\frac{{\sin}({z}/{D})}{(z/D)},
\end{equation}
which restricts the sampled velocity field to
wavenumbers satisfying $|k_z|\leq \pi/D$ and filters out all fluctuations for scales smaller than $D$. Although this vertical filtering has little influence on the large turbulent motions considered in previous applications, it becomes necessary when predicting power fluctuations over scales approaching the rotor diameter.

Taking the Fourier transform of $G(\boldsymbol{x})$ in all three
spatial directions, the transfer function magnitude for a collection of $N$ wind turbines located at $(x_i,y_i)$ for $(i=1,..N)$  is given by:
\begin{equation}
\begin{aligned}
|\hat{G}(\boldsymbol{k})|^2
={}&
\mathrm{H}\!\left(\frac{\pi}{D}-|k_z|\right)
\left(\frac{\sin(k_yD/2)}{(k_yD/2)}\right)^2
\\
&\times
\sum_{i=1}^{N} \sum_{j=1}^{N}
\cos\!\left[
k_x(x_i-x_j) +k_y (y_i-y_j)
\right].
\end{aligned}
\label{eq:G_transfer}
\end{equation}
The  horizontal direction sampling (Eq.~\ref{eq:G_xy}) was first introduced by \cite{Bossuyt2017} but we here also include $\hat{G}_z$ (i.e. the Heaviside function in $k_z$) to treat the vertical filtering. However, as turbulence is not homogeneous in the vertical direction, this step is a modeling approximation. 

Fig.~\ref{fig:fig2}a shows the transfer function calculated by Eq.~\ref{eq:G_transfer} for a $5\times6$ aligned wind farm array spaced evenly in the streamwise direction, $S_x$, and the spanwise direction, $S_y$.  A peak at $k_x = 2\pi/S_x$ and at $k_y = 2\pi/S_y$ are observed in the streamwise  and spanwise directions, respectively. 

\subsection*{The spatio-temporal spectral model of turbulent velocity fluctuations}  We adopt the Kraichnan-Tennekes random-sweeping model \cite{Kraichnan1964Kolmogorov,Tennekes1975Eulerian,Wilczek_narita_2012} further developed by \cite{Wilczek_Stevens_Meneveau_2015}, which expresses the spatio-temporal spectrum of the streamwise velocity fluctuations as
\begin{equation}
    E_{\text{uu}} (\boldsymbol{k},\omega) = \frac{E_{\text{uu}} (\boldsymbol{k})}{[2\pi\, \langle (\boldsymbol{V\cdot k})^2 \rangle]^{1/2}}\,\, \exp{\left[ - \frac{(\omega - {\bf k}\cdot {\bf U}_c)^2}{2\langle (\boldsymbol{V\cdot k})^2 \rangle} \right]},
    \label{eq:LRA_model}
\end{equation}
where $E_{\text{uu}}(\boldsymbol{k})$ is the spatial spectrum of the streamwise velocity fluctuations (here we take the full 3D spectrum), $U_c$ is the hub-height convection velocity, and $\boldsymbol{V}$ denotes the large-scale random-sweeping velocity. Here, the angle brackets $\langle \cdot \rangle$ represent a spatial and temporal average.
In this description, the smaller turbulent motions are advected by the combined action of the mean advection velocity $U_c$ and a slowly varying, zero-mean large-scale sweeping velocity field $\boldsymbol{V}$ that is assumed to be uncorrelated from the small-scale turbulence. 
Mean advection shifts the spectral energy toward the characteristic frequency $\omega=k_xU_c$, whereas random sweeping distributes the energy over a finite range of frequencies around this
value (Doppler shift and Doppler broadening, respectively). The model has been shown to reproduce the principal features of the spatio-temporal structure of wall-bounded turbulence throughout the logarithmic layer \cite{Wilczek_Stevens_Meneveau_2015,wilczek_height-dependence_2015}.
Related formulations have been applied to other atmospheric flows, including convective boundary layers \cite{Ali_2023} and atmospheric turbulence over ocean waves \cite{Ayala2026_madrid}. 
Also, Eq. \ref{eq:LRA_model}  neglects effects of mean shear \cite{mann1994spatial,Zhao2009spacetime}. For flows with strong mean shear, the modeling could be extended to include such effects. 

Assuming that the components of the random-sweeping velocity are uncorrelated, the variance of the frequency distribution can be written as $\langle (\boldsymbol{V\cdot k})^2 \rangle=\langle u'\rangle^2k_x^2+\langle v'\rangle^2k_y^2+\langle w'\rangle^2k_z^2$. To model the streamwise and spanwise sweeping variance, we invoke the attached-eddy model, which predicts a logarithmic variation of these   variances within the logarithmic layer
\cite{marusic_review_AEM,Marusic2013LogarithmicRegion}, and  the frequency variance consequently becomes
\begin{equation}
    \langle (\boldsymbol{V\cdot k})^2 \rangle=u_\tau^2 \left[B-A\log{\left(\frac{z}{h}\right)}\right](k_x^2+0.41k_y^2),
\end{equation}
where $u_\tau=\sqrt{\tau_{\mathrm{w}}/\rho}$ is the friction velocity,
$\tau_{\mathrm{w}}$ is the wall shear stress, and $\delta$ is the boundary-layer height. We use $A=0.965$ for the Perry--Townsend coefficient and $B=2.41$ for the additive constant, and  $\langle v'\rangle^2 \approx 0.41 \langle u'\rangle^2$ following the observations in \cite{Wilczek_Stevens_Meneveau_2015}. Although the reported values of these coefficients vary among flows \cite{Qin2025AttachedEddy}, such variations primarily modify the width of the frequency distribution rather than the underlying form of the model. Finally, we neglect the vertical random-sweeping contribution since  $\langle w'\rangle$ is known to be smaller than the other 2 components. 

It remains to specify the spatial spectrum $E_{\text{uu}}(\boldsymbol{k})$. Following \cite{Wilczek_Stevens_Meneveau_2015}, we represent it as a smooth blend between a low-wavenumber $k_x^{-1}$ contribution, $E_{\text{uu}}^{<}$, describing the anisotropic energy-containing motions when $k_x z<1$, and a high-wavenumber contribution, $E_{\text{uu}}^{>}$, describing the inertial-range motions when when $k_x z>1$:

\begin{equation}
    E_{\text{uu}} (\boldsymbol{k})= [1-\theta]\,{\color{black}E_{\text{uu}}^<(\boldsymbol{k})} + \theta\,{E_{\text{uu}}^>(\boldsymbol{k})},
\end{equation}
where $\theta = {(|{\boldsymbol{\tilde{k}}}|z)^4}/({1 + (|\boldsymbol{\tilde{k}}|z)^4})$ and $\boldsymbol{\tilde{k}}=(k_x,k_y)^{\text{T}}$. The high-wavenumber contribution is modeled by assuming isotropic turbulence obeying Kolmogorov scaling, thus relating ${E_{\text{uu}}^>(\boldsymbol{k})}=2\Phi_{\text{11}}(\boldsymbol{k})$, where $\Phi_{\text{pq}}({\boldsymbol k})=E(|\boldsymbol{k}|)/(4\pi |\boldsymbol{k}|^2)[\delta_{pq}-k_pk_q/|\boldsymbol{k}|^2]$ is the spectral energy tensor for homogeneous isotropic turbulence \cite{Pope2000Turbulent} written for the streamwise direction, i.e. $p=q=1$. Assuming a Kolmogorov spectrum $E(|\boldsymbol{k}|)=C_K \varepsilon^{2/3}|\boldsymbol{k}|^{-5/3}$, the high-wavenumber spectrum is
\begin{equation}
    {E_{\text{uu}}^>(\boldsymbol{k})}=\frac{C_K\varepsilon^{2/3}}{2\pi} \left( 1 - \frac{k_x^2}{|\boldsymbol{k}|^2}\right) |\boldsymbol{k}|^{-11/3},
    \label{eq:E_small}
\end{equation}
where $C_K = 1.6$ is the universal Kolmogorov constant \cite{Pope2000Turbulent}. The mean rate of energy dissipation is obtained from classical equilibrium log-layer scaling as $\varepsilon=u_\tau^3/(\kappa z)$, where $\kappa=0.4$ is the von K\'arm\'an constant. For the low-wavenumber contribution, we retain the horizontal-plane form inherited from the spectral model of \cite{Wilczek_Stevens_Meneveau_2015,Bossuyt2017} and neglect large-scale variation in the vertical direction, such that all energy is concentrated at $k_z=0$:
\begin{equation}
     E^{<}_{\mathrm{uu}}(\boldsymbol{k})
     =
     \frac{(18/55)\,C_K}{\kappa^{2/3}}\,z\,u_\tau^2 \,\delta(k_z)
     \left[\left(\frac{1}{h}\right)^6 + k_x^6\right]^{-1/6}.
     \label{eq:E_large_2D}
\end{equation}
The prefactor $(18/55)C_K\kappa^{-2/3}$ is obtained by matching the low- and high-wavenumber spectra at $k_xz=1$ and is consistent with the logarithmic streamwise-variance scaling observed by \cite{Marusic2013LogarithmicRegion}. Together, Eqs.~\ref{eq:LRA_model}--\ref{eq:E_large_2D} provide a fully parameterized description of the spatio-temporal velocity fluctuation spectrum in turbulent boundary layers. Figure~\ref{fig:fig2}b shows the streamwise wavenumber--frequency spectrum predicted by the model. Mean advection produces a Doppler-shifted convective ridge, while random sweeping broadens the spectral energy around this ridge. The model requires knowing the convection velocity $U_c$, friction velocity $u_\tau$, boundary-layer height $h$, and wind turbine hub-height $z_h$. These quantities are supplied by the wind-farm boundary-layer model (see below).

\subsection*{\textit Mean flow and the top-down wind-farm boundary-layer model}  
Flow properties such as the friction velocity $u_\tau$, boundary-layer thickness $h$, and convection velocity $U_c$ could be obtained from field measurements or high-fidelity simulations. From an engineering perspective, however, a fully predictive framework is considerably more useful and generalizable. The top-down description of fully developed wind-farm boundary layers introduced by \cite{calaf2010large} and further developed by \cite{Meneveau2012TopDown} provides the requisite predictions for the case of large wind farms. The top-down model replaces the spatially heterogeneous flow within a large wind farm by its horizontally averaged vertical structure. The wind-turbine region separates two approximately constant-stress logarithmic layers: a lower layer between the ground and the lower height of the turbine rotors and an upper layer above the tip of the rotors. These layers are characterized by friction velocities
$u_{\tau,\mathrm{lo}}$ and $u_{\tau,\mathrm{hi}}$, respectively. Within the rotor region, wake-generated turbulence enhances vertical momentum transport and produces a more strongly mixed mean-velocity profile.
The wind-farm loading is represented by the planform thrust coefficient $c_{\mathrm{ft}}={\pi C'_T D^2}/({4S_x S_y})$, where $C'_T$ is the turbine thrust coefficient (based on the disk-averaged velocity \cite{meyers2010AIAA}) and $S_x$ and $S_y$ are the streamwise and spanwise turbine spacings. The wake-enhanced mixing is modeled through a nondimensional wake eddy viscosity $   \nu_w^*\approx28({c_{\mathrm{ft}}}/{2})^{1/2}$.

The combined effects of turbine loading, wake-enhanced mixing, and the underlying surface roughness can be represented by an effective roughness length $z_{0,\mathrm{hi}}$ experienced by the boundary layer above the wind farm 
\begin{equation}
\begin{aligned}
z_{0,\mathrm{hi}}
={}
z_h
\left(
    1+\frac{D}{2z_h}
\right)^\beta
&
\exp\left\{
-
\left[
    \frac{c_{\mathrm{ft}}}{2\kappa^2}
    +
    \mathcal{L}_{\mathrm{lo}}^{-2}
\right]^{-1/2}
\right\},
\end{aligned}
\label{eq:effective_wf_roughness}
\end{equation}
where $\mathcal{L}_{\mathrm{lo}}
=
\ln\left(
{z_h}/{z_{0,\mathrm{lo}}}
\left(
1-{D}/{2z_h}
\right)^\beta
\right)$,  $z_{0,\mathrm{lo}}$ is the roughness
length of the underlying surface and $\beta=\nu_w^*/(1+\nu_w^*)$. The effective roughness $z_{0,\mathrm{hi}}$ accounts for the enhanced momentum extraction and turbulence generated by the turbine array affecting the   flow above the farm.
 
To determine the friction velocity above the wind farm, the inner-layer model is matched to an outer atmospheric boundary layer driven by a prescribed geostrophic velocity $U_G$. An approximate explicit inversion of the geostrophic drag law gives
\begin{equation}
    u_{\tau,\mathrm{hi}}
    =  \kappa U_G \,
        \left[ \ln\left(
            \dfrac{U_G}{f z_{0,\mathrm{hi}}}
        \right)
        -
        C^*\right]^{-1},
    \label{eq:upper_friction_velocity}
\end{equation}
where $f$ is the Coriolis parameter and $C^*\approx4.5$ is a known empirical constant \cite{Frandsen2006Analytical}. The corresponding atmospheric boundary-layer thickness is estimated
from the outer-layer matching relation as $h\approx 0.16\,{u_{\tau,\mathrm{hi}}}/{f}$. The friction velocity below the turbine region follows from continuity of the horizontally averaged velocity across the wake layer:

\begin{equation}
    u_{\tau,\mathrm{lo}}= u_{\tau,\mathrm{hi}}
   \, \ln(Z_1 Z_+^\beta)/\ln(Z_1 Z_-^\beta),\label{eq:lower_friction_velocity}
\end{equation}
where $Z_1=z_h/z_{0,\mathrm{hi}}$ and 
$Z_\pm=[1\pm (D/2z_h)]^\beta$. The mean hub-height velocity can then be evaluated from either the upper or lower logarithmic layer as

\begin{equation}
U_h=
\frac{u_{\tau,\mathrm{hi}}}{\kappa}
\ln(Z_1\,Z_+^\beta).
\label{eq:hub_velocity_topdown}
\end{equation}
All required flow parameters and then be determined based on input quantities such as the geostrophic forcing wind speed $U_G$, the wind farm's underlying soil roughness $z_{0,lo}$, the latitude that determines the Coriolis frequency $f$, and turbine properties $D$ and $z_h$.

\section*{Comparison against high-fidelity wind-farm simulations}
\label{sec:section2}

We now evaluate the complete model against a high-fidelity wind-farm simulation dataset for which all of the required input quantities are known. The comparison is performed using the JHTDB-wind dataset introduced by \cite{Zhu2025JHTDBWind}. The database contains time-resolved large-eddy simulation fields and turbine-response signals (aerodynamic power) for a 60-turbine wind farm operating within a conventionally neutral atmospheric boundary layer. The simulated wind farm consists of $10$ streamwise rows and $6$ spanwise columns. We focus on the final five rows, corresponding to a total of $30$ turbines, which are taken to represent the approximately fully developed region. The turbine diameter and hub height are $D=126$~m and $z_h=90$~m, respectively, while the streamwise and spanwise spacings are $S_x=7$ and $S_y=5$. These quantities fully specify the spatial sampling function for the portion of the farm considered in the comparison.

To specify flow properties, we use the atmospheric conditions prescribed in JHTDB-wind, i.e.  $f=10^{-4}$~s$^{-1}$ (corresponding to a representative mid-latitude location), $U_G=15$~m~s$^{-1}$, and
an underlying  surface roughness length of $z_{0,\mathrm{lo}}=0.1$~m. 
The convection velocity entering the spatio-temporal spectrum is assumed to equal the mean velocity, $U_c=U_h$ (this is the simplest choice but it is known that fluctuation advection velocities often differ from the mean velocity \cite{DELÁLAMO_JIMÉNEZ_2009}, hence further model improvements are possible). The mean velocity $U_h$ is predicted by the top-down model. The friction velocity entering the turbulence-spectrum parameterization is taken as $u_\tau=u_{\tau,\mathrm{lo}}$. The mean disk-averaged velocity required in $C_2$ is estimated from actuator-disk theory as $\langle\overline{u}\rangle_D=U_c(1-a)$, for which we adopt the typical induction factor $a=0.25$ \cite{calaf2010large}. For the power-fluctuation model, we use $C_P'=0.715$ and $C_T'=0.562$ (see supplementary material on how these coefficients are determined in our analysis).  Table~\ref{tab:tab1} compares key flow quantities extracted from the database with their corresponding predictions from the top-down model. Overall, the model predictions show reasonable agreement with the database estimates  across all quantities considered.

\begin{table}[t!]
\centering
\caption{Comparison of flow quantities between the reference data and the top-down model.}
\label{tab:tab1}
\begin{tabular}{lcccc}
 & $u_\tau$ (m/s) & $\langle \overline{u}\rangle_D$ (m/s) & $U_c$ (m/s)  \\
\midrule
JHTDB-Wind & 0.607 & 6.873 & 9.700  \\
Model & 0.604 & 6.765 & 9.020 \\
\bottomrule
\end{tabular}
\end{table}

\begin{figure*}[t!]
    \centering
    \includegraphics[width=0.95\linewidth]{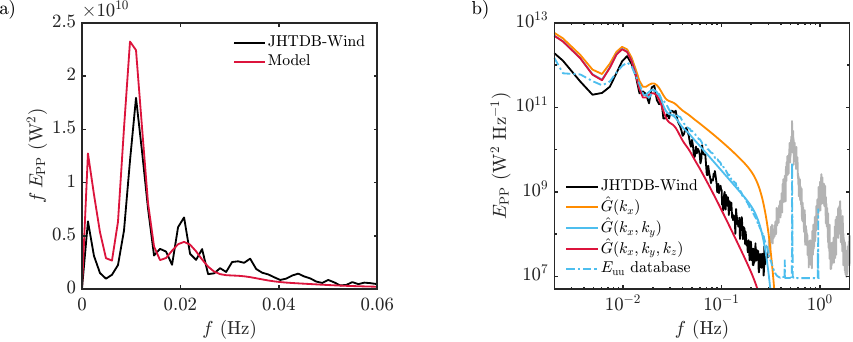}
    \caption{a) Comparison of the premultiplied power spectrum of the aggregate power fluctuations from 30 turbines, obtained from the JHTDB-Wind database and predicted by the spectral model. b) Aggregate power spectrum predicted using the full three-dimensional sampling function $\hat{G}(k_x,k_y,k_z)$, the horizontal-only $\hat{G}(k_x,k_y)$ and the streamwise-only $\hat{G}(k_x)$, all with the modeled $E_{\mathrm{uu}}$. Dash-dotted line uses the  LES database-derived $E_{\mathrm{uu}}$ instead of the modeled one for comparison. }
    \label{fig:fig3}
\end{figure*}

Figure~\ref{fig:fig3} compares the power spectral density obtained from JHTDB-wind with the predictions of the spectral model using both linear and logarithmic axes. As can be seen, the model (red line in both panels) reproduces the principal features of the data remarkably well for frequencies below 0.3 Hz. In particular, it captures the spectral peak near $f\approx0.01$~Hz, which is associated with the convective travel time between successive turbine rows ($f=U_c/S_x=9.7/882=0.01$) and the resulting coherence among their power signals when added together.  A small shift in the predicted spectrum is also observed, which can be attributed in part to the lower convection velocity estimated by the top-down model.  The model also reproduces the decay across the inertial-frequency range between 0.15 and 0.3 Hz. In the present formulation, this attenuation arises from spatial averaging of turbulent fluctuations over the finite rotor area, with the added vertical sampling playing an essential role in filtering turbulent motions whose scales approach the rotor diameter.  At higher frequencies,  the model does not reproduce the narrow high-frequency peaks associated with blade rotation and blade-passing harmonics since these motions have been filtered out  through disk averaging, which eliminates rotational flow sampling and blade-resolved aerodynamic effects. Such peaks can be modeled using more detailed approaches based on rotationally sampled turbulence and blade-passing-frequency harmonics \cite{Deskos2020}. The good agreement within the frequency range of interest (below 0.3 Hz) is notable given that the prediction is obtained entirely from wind-farm geometry, turbine operating parameters, and large-scale atmospheric inputs, without calibration from the wind farm data.

\section*{Model sensitivity and predictive scope}
\label{sec:section3}

\subsection*{Effects of spatial sampling and turbulence spectrum representation} We  isolate the role of the spatial sampling function by comparing three representations: the three-dimensional transfer function $\hat{G}(k_x,k_y,k_z)$ introduced in Eq.~\ref{eq:G_transfer}, the two-dimensional formulation $\hat{G}(k_x,k_y)$ of \cite{Bossuyt2017} given by Eq.~\ref{eq:G_xy}, and a one-dimensional function $\hat{G}(k_x)$ that retains only the discrete streamwise turbine locations. For the latter, we adopt $G(x)=\sum_{i=1}^{N}\delta(x-x_i)$ as a proxy for streamwise point sampling without spanwise or vertical rotor averaging. Figure~\ref{fig:fig3}b shows the resulting aggregate power spectra for the three cases.  Increasing the dimensionality of the sampling function progressively strengthens the spectral decay in the inertial range, while leaving the largest scales nearly unchanged. This behavior reflects the fact that spanwise and vertical rotor averaging primarily attenuate turbulent motions with wavelengths comparable to or smaller than the rotor diameter. The full three-dimensional sampling function is therefore required to represent the finite-area filtering of the rotor and recover the observed inertial-range attenuation (see also \cite{Druault2022Spatial}).

We next isolate the influence of the turbulence spectrum $E_{\mathrm{uu}}$. For this comparison, we use the spatio-temporal spectrum extracted directly from the JHTDB-wind data at hub height. Since this spectrum was computed on a single horizontal plane, it is resolved only in $(k_x,k_y,\omega)$ and cannot be combined consistently with the full three-dimensional transfer function. Figure~\ref{fig:fig3}b (blue lines) compares the power spectra obtained using the modeled and database-obtained turbulence spectra. The database spectrum yields improved agreement  at the largest scales, particularly near the row-to-row advection peak, which is expected as it contains more accurate convection velocity and spatio-temporal organization from the data itself. It also produces narrow high-frequency peaks associated with blade rotation, whose signatures are embedded in the simulated flow field but absent from the present model. Apart from these minor differences, the model reproduces the power spectrum with similar accuracy as that based on the actual turbulence spectrum. 

\subsection*{Layout dependence of the aggregate power spectrum} 
\begin{figure*}[b!]
\centering
\includegraphics[width=\linewidth]{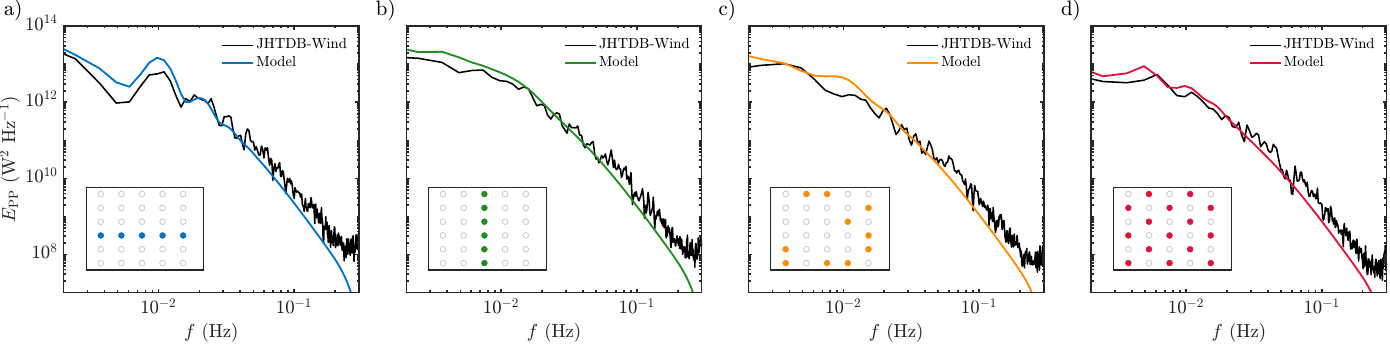}
\caption{a) A single spanwise column ($N=5$), b) a single streamwise row ($N=6$), c) ten randomly selected turbines ($N=10$), and d) a staggered arrangement in which alternate rows sample alternate columns ($N=15$). Insets show the turbines retained in each layout (filled symbols) within the full $5\times6$ array.}
\label{fig:fi5}
\end{figure*}

We now characterize how wind-farm layout shapes the aggregate power spectrum. To enable direct comparison with JHTDB-wind, four turbine subsets are selected from the $5\times6$ array: a single streamwise column, a single spanwise row, 10 randomly distributed turbines, and a staggered subarray. Figure~S5 in the Supplementary Material shows the sampling transfer function $\hat{G}(k_x,k_y)$ used for each sub-array configuration. Figure~\ref{fig:fi5} compares the modeled and reference spectra for each configuration. For the streamwise column (Fig.~\ref{fig:fi5}a), the model captures the spectral peak associated with advection between successive turbine rows. For the spanwise row and randomly selected turbines (Fig.~\ref{fig:fi5}b,c), it reproduces the reference spectra reasonably well across the modeled frequency range without clear peaks, although the random case (Fig. \ref{fig:fi5}c) displays some minor differences in low frequency undulations. In the staggered configuration (Fig.~\ref{fig:fi5}d), the increased streamwise separation between successive selected turbines shifts the advection peak toward lower frequencies relative to the aligned case. The model captures this layout-induced shift, with the remaining frequency offset arising primarily from the predicted convection velocity. These comparisons show that the framework (considering wind farms as sensor arrays of turbulent boundary layers) successfully predicts spectral features of wind farm aggregate power fluctuations.

\subsection*{Exploring the model for various atmospheric forcings and surface roughness} We next examine the predictive scope of the model under different atmospheric conditions while retaining the same $5\times6$ turbine layout. To isolate the effects of the atmospheric inputs, we vary $z_{0,\mathrm{lo}}$, $U_G$ and  $h$, while holding $C'_T$ and $C'_P$ fixed.  In practice, these coefficients may change with the turbine operating state, particularly when turbines are operating near or above rated conditions. Four conditions are considered. Case~1 is the baseline JHTDB-wind configuration. Case~2 represents an offshore-like smooth surface, with the same geostrophic forcing as Case~1 but $z_{0,\mathrm{lo}}=10^{-4}$~m. Case~3 represents a lower-wind onshore condition. Case~4 represents a shallow onshore boundary layer over a forest-like rough surface, with a larger streamwise separation between turbine rows. Cases~3 and 4 are introduced to illustrate sensitivity of the model to reduced boundary-layer depth. The flow quantities for cases 1-4 are summarized in Table~\ref{tab:tab2}.

\begin{table}[t!]
\centering
\caption{Input parameters and top-down model predictions for the four atmospheric and layout cases.}
\label{tab:tab2}
\begin{tabular}{cccccccc}
Case
& $U_G$
& $z_{0,\mathrm{lo}}$
& $S_x/D$
& $h$
& $u_\tau$
& $U_c$ \\
& (m/s)
& (m)
& ($-$)
& (m) 
& (m/s)
& (m/s) \\
\midrule
1 & 15 & $10^{-1}$ & 7 & 1500 & 0.60  & 9.02   \\
2 & 15 & $10^{-4}$ & 7 & 1300  & 0.30  & 9.71   \\
3 & 5  & $10^{-1}$ & 7 & 600 & 0.24 & 3.63   \\
4 & 8  & $1$ & 15 &  300 & 0.53   & 4.85   \\
\bottomrule
\end{tabular}
\end{table}

\begin{figure}[t!]
\centering
\includegraphics[width=0.75\linewidth]{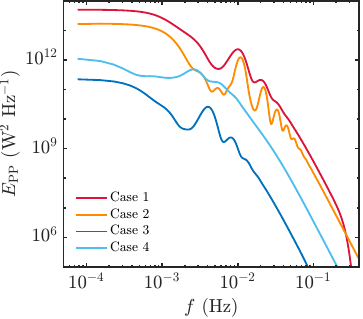}
\caption{Aggregate power spectra predicted by the fully parameterized model for three atmospheric conditions.}
\label{fig:fig6}
\end{figure}

Figure~\ref{fig:fig6} compares the predicted aggregate power spectra for  cases 1--4 in Table~\ref{tab:tab2}. Comparing cases~1 and 2 isolates the effect of surface roughness at fixed geostrophic forcing. The smoother offshore-like surface produces a similar convection velocity but substantially smaller friction velocity and velocity variance. Consequently, the principal advection peak remains at approximately the same frequency, while the broadband spectral energy decreases by roughly a factor of five. The lower random-sweeping variance also produces a narrower spatio-temporal frequency distribution, making the row-to-row peak and its harmonics more pronounced.

Comparing Case ~1 to Case 3 isolates the effect of reducing the geostrophic forcing over the same surface. The lower convection velocity shifts the row-to-row spectral peaks toward lower frequencies, consistent with their characteristic scaling of $f\sim U_c/S_x$. The spectral magnitude also decreases by nearly two orders of magnitude because both the disk-averaged velocity and the friction velocity are reduced. 
Case~4 includes the combined effects of a rough surface, a shallow boundary layer, and increased streamwise turbine spacing. Although its convection velocity is larger than that of Case~3, increasing the row spacing from $S_x/D=7$ to $S_x/D=15$ shifts the fundamental row-to-row frequency. At the same time, the larger friction velocity and streamwise velocity variance increase the random-sweeping broadening. The resulting overlap between adjacent layout harmonics produces a broad low-frequency shoulder rather than a sequence of sharply resolved peaks. 

\section*{Conclusions}

We have developed a fully predictive model for the aggregate power-fluctuation spectrum of a wind farm operating within a fully developed, conventionally neutral boundary layer. The framework combines three-dimensional sampling by the turbine array, a model for the spatio-temporal turbulence spectrum, and a top-down description of the wind-farm boundary layer, thereby connecting atmospheric forcing, turbine properties, and farm geometry directly to the power spectrum. Comparison with data from a high-fidelity database (JHTDB-wind) shows that the model captures the principal spectral features of the data, including the peak associated with advection between turbine rows and the attenuation of inertial-range fluctuations by finite-rotor averaging.  Additional analyses clarify the physical roles of the model components. These results demonstrate how fundamental fluid-dynamics and boundary layer turbulence theory can be leveraged to provide accurate predictions of the temporal spectrum of wind power fluctuations, a quantity of direct importance to efforts aiming to increase wind energy contributions to  electricity grids. 

\matmethods{

\subsection*{Details of the JHTDB-Wind database}The analysis uses the JHTDB-wind dataset \cite{Zhu2025JHTDBWind}  containing one hour of large-eddy simulation data for a conventionally neutral atmospheric
boundary layer interacting with 60 NREL 5-MW reference turbines
arranged in 10 streamwise rows and 6 spanwise columns. The turbines
have rotor diameter $D=126$~m and hub height $z_h=90$~m, with
streamwise and spanwise spacings of $7D$ and $5D$, respectively.
The boundary layer is driven by a geostrophic wind of magnitude
$G=15$~m~s$^{-1}$ over a surface with roughness length
$z_0=0.1$~m, and the mean hub-height wind is aligned with the
streamwise direction. The simulation was performed using LESGO \cite{lesgo} with a concurrent-precursor inflow and a filtered actuator-line representation of the turbines.  The stored domain spans $22.932\times3.78\times2.0$~km$^3$. Flow-field quantities are stored on a $1248\times192\times400$ grid every $0.5$~s, whereas turbine-level power, thrust, and rotor-speed signals are stored every $\Delta t=0.025$~s, yielding 144,000 temporal samples per turbine. The model evaluation uses the 30 turbines in the final five rows, where the wind-farm boundary layer is approximately fully developed.
Figure~S4a in the Supplementary Material shows a schematic representation of the large wind farm during 1-hour conventionally neutral atmospheric conditions database and Fig~S4b shows a snapshot of the flow field at hub height level.
}

\newpage
\showmatmethods{}

\dataavail{The data used in this study can be obtained from the Johns Hopkins Turbulence Databases (JHTDB; \hyperlink{https://turbulence.idies.jhu.edu/datasets/windfarms/neutralWindfarm}{https://turbulence.idies.jhu.edu/datasets/windfarms/neutralWindfarm})}

\acknow{This work was supported by the National Science Foundation and the Department of Energy (via NSF grant CBET-2401013).}

\showacknow{} 

\bibsplit[25]

\end{document}